\documentclass[twocolumn,showpacs,superscriptaddress,10pt]{revtex4-2}
\usepackage{amsmath,amssymb,graphicx, subfigure, hyperref, braket, float}
\usepackage{blindtext}

\counterwithout{figure}{section}
\usepackage[export]{adjustbox}
\graphicspath{{figure/}}
\usepackage{graphicx, xcolor}
\usepackage[margin=.8in,text={10in,10.5in},centering]{geometry}
\usepackage{graphicx}
\usepackage{hyperref}
\usepackage{lipsum}
\usepackage{caption}
\usepackage{tikz}
\usepackage{needspace} 
\usepackage{bbold}
\usepackage{pgf}
\usepackage{ragged2e} 
\usepackage{booktabs}

\allowdisplaybreaks
\hypersetup{ colorlinks=true,linkcolor=red,citecolor=blue,urlcolor=magenta}
\begin{document}
	
\title{Nonreciprocal Macroscopic Entanglement through Magnon Squeezing in a Cavity Magnomechanics}

\author{Ziyad Imara} \email{imara.ziyad@etu.uae.ac.com}
\address{Laboratory of R\&D in Engineering Sciences, Faculty of Sciences and Techniques Al-Hoceima, Abdelmalek Essaadi University, Tetouan, Morocco}

\author{Khadija El Anouz}\email{kelanouz@uae.ac.ma}
\address{Laboratory of R\&D in Engineering Sciences, Faculty of Sciences and Techniques Al-Hoceima, Abdelmalek Essaadi University, Tetouan, Morocco}

\author{İlkay Demir}\email{idemir@cumhuriyet.edu.tr}
\address{Department of Nanotechnology Engineering, Sivas Cumhuriyet University, 58140 Sivas, Türkiye}
\address{Sivas Cumhuriyet University Nanophotonics Application and Research Center-CÜNAM, 58140 Sivas, Türkiye}

\author{Abderrahim El Allati}\email{eabderrahim@uae.ac.ma}
\address{Laboratory of R\&D in Engineering Sciences, Faculty of Sciences and Techniques Al-Hoceima, Abdelmalek Essaadi University, Tetouan, Morocco}

\begin{abstract}
Cavity magnomechanics has opened a new frontier in quantum electrodynamics, yielding several significant theoretical and experimental results. In this paper, we propose a different theoretical mechanism to achieve nonreciprocal macroscopic entanglement among magnons, photons, and phonons, based on magnon squeezing. 
Specifically, reversing the squeezing phase, namely $\theta\to\theta+\pi$ reverses the frequency shift and the effective dissipation rate simultaneously, producing two experimentally distinct configurations that enable nonreciprocal entanglement.
Indeed, in contrast to conventional approaches that control only frequency shifts, we show how precise control of the amplitude and phase of the squeezed mode allows us to obtain a tunable nonreciprocity of entanglement. The magnons resulting from the collective motion of the spin in a macroscopic ferrimagnet become coupled to the microwave photons via magnetic dipole interaction and to the phonons via magnetostrictive interaction. Moreover, we show that the proposed scheme achieves ideal nonreciprocity, which can be optimized by cavity-magnon coupling and bath temperature control. Finally, by using the parameters that are experimentally feasible with current technologies, this work provides promising perspectives for hybrid magnon-based quantum technologies.
\end{abstract}

\vspace{2cm}
\maketitle
\section{Introduction}
\label{sec1}

The field of Magnonomechanics (MM) emerges as a highly promising platform for exploring macroscopic quantum effects \cite{A01,A2,A4}. It consists of investigating the interaction between spin waves (magnons) and mechanical vibrations (phonons) via magnetostrictive coupling \cite{A1}.  These systems mainly exploit yttrium-iron garnet (YIG), a material exhibiting remarkable magnetic properties that are proving key to advanced magneto-optical applications. In fact, across various geometries, including sphere \cite{A3,S1,S2}, thin film \cite{A5} or bridge \cite{A6,A7,XX1}, the YIG system enables efficient magnon-phonon coupling through magnetostrictive force. Remarkably, these systems have theoretically shown the ability to achieve photon-magnon-phonon entanglement, paving the way for breakthrough applications in quantum technologies, notably in information storage, ultrasensitive detection, quantum transduction, and hybrid quantum networks \cite{A8,A9,B0,B1}. Magnons in YIG can achieve strong coupling with microwave photons in a high-quality cavity \cite{D6,D7}. This gives rise to magnon–cavity polaritons through the collective excitations of a large number of spins \cite{B3}. Moreover, squeezed magnons are a powerful tool for improving the performance of cavity magnomechanics systems, enhancing quantum entanglement, supporting nonlinearity to be amplified and cooling of the ground state to be improved \cite{B4,B04}. Various methods can be used to achieve magnon squeezing, including exploiting the intrinsic nonlinearity of the magnetostriction \cite{B5}, reservoir-engineered cavity magnomechanics \cite{B7}, two-tone microwave driving \cite{B8}, and other techniques \cite{B9,C0,C1,XXXX1}.

Notably, MM systems can achieve nonreciprocal entanglement between photons and magnons by breaking time-reversal symmetry. This phenomenon represents directional asymmetry where quantum correlations exist for one configuration but are suppressed when that configuration is reversed. The concept was first introduced in rotating optomechanical resonators, where mechanical rotation induces opposite frequency shifts for clockwise and counterclockwise propagating modes \cite{C2,CX2}. Recently, nonreciprocal entanglement in MM systems has been demonstrated by exploiting effects that induce frequency shifts in magnon resonances, analogous to the optomechanical Sagnac effect \cite{C9,C3,C4,C44}. Several approaches have been proposed, including the Kerr effect that produces state-dependent frequency shifts \cite{C6,CCC5,C8,ZZia}. For instance, Chen \textit{et al.} exploited the Kerr nonlinearity, where the sign of the magnon frequency can be reversed by reorienting the magnetic field along different crystallographic axes of the YIG sphere \cite{C8}. Other methods include spinning the MM cavity \cite{C6}, exploiting chiral coupling \cite{C9}, and the Barnett effect, where rotation induces opposite magnetization polarities depending on the rotation direction \cite{D1,D2}. However, all these approaches share a common feature in controlling only the frequency shift of the magnon mode.

In this paper, we propose a fundamentally distinct approach to achieve nonreciprocal entanglement based on magnon squeezing. Our key innovation is dual control, where the squeezing phase $\theta$ manipulates simultaneously the frequency shift and the dissipation rate of the magnon mode. Physically, $\theta$ corresponds to the phase of the parametric drive generating the magnon squeezing, in direct analogy with the treatment of parametric squeezing in optomechanical systems \cite{op1}. 
Reversing the squeezing phase under the transformation $\theta \to \theta + \pi$ produces two experimentally distinct configurations with opposite frequency shifts and dissipation rates, analogous to magnetic axis reversal in the Kerr effect \cite{C8} and rotation reversal in the Barnett effect \cite{D1}, enabling ideal nonreciprocity across broader parameter regimes than existing single-parameter methods. This approach provides in situ control compared to the conventional techniques, using achievable standard squeezed magnons in cavity magnomechanics. We demonstrate how squeezing amplitude, phase, temperature, and cavity-magnon coupling enable robust nonreciprocal entanglement in a stable regime and verify the presence of genuine magnon squeezing for all configurations used in our study.

The present paper is structured as follows. In Sec.~\ref{sec2}, we present the theoretical model of the system, where we derive the linearized quantum Langevin equations of the system. In Sec.~\ref{sec3}, we perform numerical simulations using experimentally accessible parameters to show the nonreciprocity of bipartite and tripartite entanglement, followed by a detailed discussion of the results. Finally, Sec.~\ref{sec4} summarizes the main results of our paper.

\vspace{-0.9em}

\section{MODEL AND DYNAMICAL HAMILTONIAN}
\label{sec2}

\begin{figure}[t]
	\centering
	\subfigure{\includegraphics[scale=0.5]{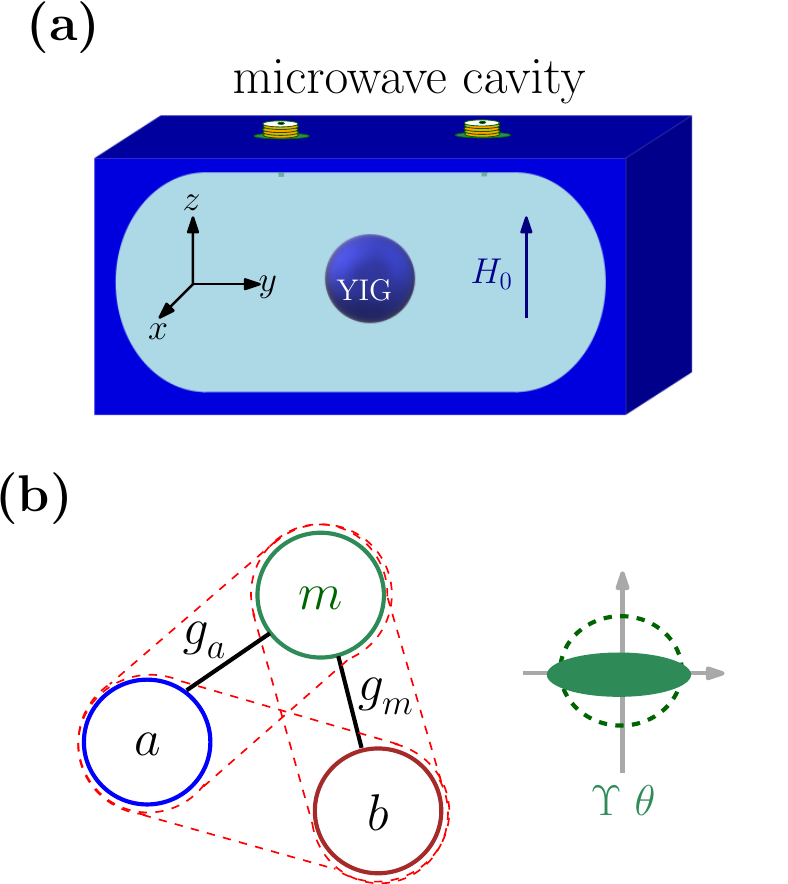}}
	\captionsetup{justification=RaggedRight, singlelinecheck=false}
	\caption{{\small (a) Schematic of the cavity magnomechanics system with a YIG sphere in a polarizing magnetic field. A polished YIG sphere is placed inside a microwave cavity close to the maximum magnetic field region of the cavity mode. (b) Schematic of all possible bipartite entanglement and couplings model with photon-magnon ($g_a$) and magnon-phonon ($g_m$) couplings. Modes $a$, $m$ and $b$, represent cavity, magnon, and phonon, respectively. Green ellipse indicates magnon mode squeezing with parameter $\Upsilon$ and phase $\theta$, enabling nonreciprocal entanglement.}}
	\label{fig:1}
\end{figure}

We consider a hybrid cavity magnomechanics system consisting of a $250\mu$m-diameter YIG sphere positioned at an amplitude belly of a microwave cavity \cite{A1,A2}, as shown in Fig.~\ref{fig:1}. 
This system couples microwave photons from the cavity, magnons and phonons. The magnons interact with the phonons via magnetostrictive coupling, which induces geometrical deformation in the YIG sphere and modifies its vibrational modes \cite{D3,D4}. Thus, it establishes a mutual interaction in which the mechanical vibrations of the sphere also modify the properties of the magnons and photons in the cavity \cite{D5}.
It is worth mentioning that the radiation pressure effects can be neglected when the YIG sphere is much smaller than the microwave wavelength.
The Hamiltonian of the system in the rotating frame at the drive frequency $\omega_{0}$ (see the \textbf{Appendix A} for more details) is given as
\begin{eqnarray}\label{Hamil}
	\label{1}
		\mathbf{H}&=&\Delta_a a^\dagger a +\Delta_{m} m^\dagger m+\frac{\omega_{b}}{2}(q^2+p^2)\nonumber\\ 
		&+& g_{_m}m^\dagger mq + g_{_a} (a^\dagger m+ a m^\dagger)\nonumber\\
		 &+& i[\Omega_0m^\dagger + \frac{\Upsilon}{2}m^{\dagger2} e^{i\theta} -\text{H.c.}],~~
\end{eqnarray}
where $a (a^\dagger)$ and $m(m^\dagger)$ are the annihilation (creation) operators of the cavity mode and the magnon mode, respectively. Furthermore, $\mathcal{O}\equiv a,m$ satisfies $[\mathcal{O},\mathcal{O}^\dagger]=\mathbf{1}$.
Here, ${\Delta}_{a} = \omega_{a} - \omega_0$, ${\Delta}_{m} = \omega_{m} - \omega_0$ are the detunings of the microwave cavity mode and magnon mode, respectively.
For the mechanical vibration modes, $q=(b^\dagger+b)/{\sqrt{2}}$ and $p={i}(b^\dagger-b)/{\sqrt{2}}$ are the dimensionless quadratures of position and momentum, where $b$ ($b^\dagger$) is the annihilation (creation) operator of phonon mode, satisfying $[q,p]=i$. Here, $\omega_{a}$ ($\omega_{b}$) denotes the frequency of the microwave cavity (the phonon) mode. Moreover, to excite and generate the magnon mode, the YIG spheres are usually placed in a uniformly biased magnetic field $H_{_0}$, while a microwave field with magnetic component $H_{_d}$ is perpendicularly applied to the polarization field \cite{D6,D7,D8}.
In addition, $\omega_{m}$ is the magnon mode frequency given by means of the external bias magnetic field and the gyromagnetic ratio $\gamma$ as $\omega_{m}=\gamma H_{_0}$, where $\gamma/2\pi= 28$ GHz/T. 
$g_{_a}$ is the cavity-magnon coupling strength that can achieve the strong coupling regime for $g_{_a}>\kappa_{_a},\kappa_{m}$, where $\kappa_a$ and $\kappa_m$ are the cavity and magnon decay rates, respectively \cite{D6,D7}. $g_{_m}$ is the magnon-phonon coupling strength and is typically small \cite{A1,A01,A2,D9}. Moreover, under the assumption of low-lying excitations, we have $\braket{m^\dagger m}\ll2N_0s$, where $s=5/2$ is the spin number of the ground state Fe$^{3+}$ ion in YIG. Here, $N_0=\rho V$ is the total number of spins, with $\rho=4.22\times10^{27}m^{-3}$ being the spin density and $V$ the volume of the YIG-sphere. Note that the magnon mode can be driven by a microwave field using, e.g., a loop antenna \cite{J1}.
The Rabi frequency associated with the microwave drive is given by $\Omega_0=\frac{\sqrt{5}}{4}\gamma\sqrt{N_0}H_{d}$.
The last term in the Hamiltonian (\ref{Hamil}), $\Upsilon/2 (m^{\dagger2} e^{i\theta}-\text{H.c.})$, denotes the magnon squeezing interaction, characterized by the squeezing amplitude $\Upsilon$ and phase $\theta$, where $\theta$ corresponds to the phase of the parametric drive generating the magnon squeezing \cite{op1}.
As reported in Ref.~\cite{B04}, this magnon squeezing can be achieved by various experimental approaches, e.g., applying two-tone microwave fields to drive the magnon mode \cite{B8}, exploiting the magnetic anisotropy of ferromagnets \cite{E0,E000}, or using magnetostriction nonlinearity \cite{B4}.

Following the standard approach \cite{Rev02,Rev2,A2,Rev3}, we incorporate dissipation and input noise terms for each mode to obtain the set of quantum Langevin equations (QLEs) for the system as follows (see \textbf{Appendix A} for details):
\begin{eqnarray}
	\label{2}
	\dot{a} &= &-(i\Delta_a + \kappa_{_a}) a - i g_{_a} m + \sqrt{2\kappa_a} a^\text{in}, \nonumber\\
	\dot{m} &= &-(i\Delta_m +\kappa_m) m - i g_{_a} a + \Omega_0 - i g_{m} m q + \Upsilon m^\dagger e^{i\theta} \nonumber \\ &+& \sqrt{2\kappa_m} m^\text{in}, \nonumber\\
	\dot{q} &= &\omega_b p~~,~~	\dot{p} = -\omega_b q - \gamma_b p - g_{_m} m^\dagger m + \xi, 
\end{eqnarray}
where $\gamma_{b}$ is the damping rate of the phonon mode. 
Moreover, $o^{in}$ is the input noise operator of mode $o=(a,m)$, which has zero mean \cite{E1}. In fact, it is characterized by the following correlation functions
\begin{eqnarray}
		\label{T1}
	\braket{o^{in\dagger}(t)o^{in}(t') } &=& \bar{n}_o \delta(t - t'),\nonumber\\
	\braket{o^{in}(t)o^{in\dagger}(t')} &=& (\bar{n}_o + 1) \delta(t - t').
\end{eqnarray}
Moreover, $\xi$ is the Brownian noise operator of the mechanical mode with zero mean value $\braket{\xi} = 0$. In the Markov approximation, this correlation function for a mechanical mode with high quality factor $\mathcal{Q} \gg 1$ takes the form \cite{E2}
\begin{equation}
	\label{3}
	\braket{\xi(t)\xi(t') + \xi(t')\xi(t)} / 2 = \gamma_{b} (2\bar{n}_{b} + 1) \delta(t - t'),
\end{equation}
where $\bar{n}_o = (\exp({\hbar\omega_o}/(k_B T)) - 1)^{-1}$ denoting the mean thermal occupation number of the $o$-mode ($o = a, m, b$) at bath temperature $T$ and the Boltzmann constant $k_B$ \cite{E3}.

Using strong coherent driving fields on the cavity and magnon modes leads to large steady-state amplitudes $|a_{s}|, |m_{s}|\gg1$. Thus, one should  linearize the QLEs in Eq.~(\ref{2}) around the steady-state value $\mathcal{X}_s$ as $\mathcal{X} = \mathcal{X}_s + \tilde{\mathcal{X}}$, where $\mathcal{X} = a, m, q, p$, and neglect the second-order fluctuation terms. The linearized QLEs describe the quantum fluctuations of the system using the quadrature fluctuation operators ($\tilde{I}_o,\tilde{J}_o$), where  $\tilde{I}_o={ (\tilde{o} +\tilde{o}^\dagger)/}{\sqrt{2}}$, $\tilde{J}_o=i{(\tilde{o}^\dagger-\tilde{o})/}{\sqrt{2}}$ with $o = a, m$, and can be written in the following matrix form 
\begin{equation}
	\label{5}
	\dot{\Phi}(t)=\Gamma\Phi(t)+\mathcal{N}(t),
\end{equation}
where $\Phi(t)=[\tilde{I}_a,\tilde{J}_a,\tilde{I}_m,\tilde{J}_m,\tilde{q},\tilde{p}]^T$ and $\mathcal{N}(t)=[\sqrt{2\kappa_a}\tilde{I}_{a}^\text{in},\sqrt{2\kappa_a}\tilde{J}_{a}^\text{in},\sqrt{2\kappa_m}\tilde{I}_{m}^\text{in},\sqrt{2\kappa_m}\tilde{J}_{m}^\text{in},0, \xi]^T$ is the vector of input noises, where  $\tilde{I}_o^{\text{in}}={ (\tilde{o}^{\text{in}} +\tilde{o}^{\dagger\text{in}})/}{\sqrt{2}}$, $\tilde{J}_o^{\text{in}}=i{(\tilde{o}^{\dagger\text{in}}-\tilde{o}^{\text{in}})/}{\sqrt{2}}$.
Moreover, $\Gamma$ is the drift matrix, given by 
\begin{equation}
	\label{6}
	\Gamma =\begin{pmatrix}
		-\kappa_{_a} & {\Delta}_a &0 & g_a & 0 & 0  \\
		-{\Delta}_a & -\kappa_{_a} &-g_a & 0 & 0 & 0  \\
		0 & g_a &-\kappa_m + \kappa_{_{\theta}} & \bar{\Delta}_m + {\Delta}_{\theta} & -G_{m} & 0  \\
		-g_a & 0 &-\bar{\Delta}_m + {\Delta}_{\theta}  & -\kappa_m - \kappa_{_{\theta}} & 0 & 0  \\
		0 & 0 &	0 & 0 & 0 & \omega_{b}  \\
		0 & 0 &	0 & G_{m} & -\omega_{_b} & -\gamma_{b}  \\
	\end{pmatrix},
\end{equation}
where $\bar{\Delta}_{m}={\Delta}_{m}+ g_m q_{_s}$ is the effective magnon detuning including the frequency shift due to magnomechanical interaction, and ${\Delta}_{{\theta}}= \Upsilon \sin\theta$ is the frequency detuning induced by the magnon squeezing.
Typically, these frequency shifts are small, i.e., $|\bar{\Delta}_{m}-{\Delta}_{m}|\ll{\Delta}_{m}$ \cite{A1,E4}.
Besides, $\kappa_{_{\theta}}= \Upsilon \cos\theta$ is the effective decay rate of the magnon mode.
$G_{m}=i\sqrt{2}g_{_m}{m}_s$ is the enhanced magnomechanical coupling strength through coherent driving, where ${q}_s=-ig_m|{m}_s|^2/\omega_{_b}^2$. While, ${m}_s$ is expressed as
\begin{eqnarray}
	\label{4}	
	m_{s}&=&\braket{m}=\frac{\varkappa_{_-}(i{\Delta}_{a}+\kappa_{a})+\Upsilon({\Delta}_{a}^2+\kappa_{a}^2)e^{i\theta}}{\varkappa_{_-}\varkappa_{_+}-\Upsilon^2({\Delta}_{a}^2+\kappa_{a}^2)}\Omega_0,
\end{eqnarray}
where $\varkappa_{_\pm}=(\kappa_a\pm i\Delta_a)(\kappa_m\pm i\bar{\Delta}_m)+g_a^2$. Hence, $m_{s}$ in Eq.(\ref{4}) can be expressed as the following form
\begin{eqnarray}
	\label{04}	
	m_{s}&=&\frac{\Upsilon e^{i\theta}+i\eta}{\eta^2-\Upsilon^2}\Omega_0,
\end{eqnarray}
where $\eta=(g_{_a}^2/\Delta_a)-\bar{\Delta}_m$, when $|{\Delta}_{a}|,|\bar{\Delta}_{m}|\gg \kappa_{a},\kappa_{m}$.\\

The drift matrix in Eq.~(\ref{5}) is given under the condition $|{\Delta}_{a}|,|\bar{\Delta}_{m}|\simeq\omega_{b}\gg \kappa_{a},\kappa_{m}$, which is the optimal condition for quantum correlations in the system \cite{A2}. As a result of the linearized dynamics and the Gaussian nature of the quantum noise, its state can be fully described as the steady state of the quantum fluctuations of the system. Indeed, it is characterized by the covariance matrix (CV) $\mathcal{V}$ with components $	\mathcal{V}_{ij}(t)=\braket{\Phi_{i}(t)\Phi_{j}(t')+\Phi_{j}(t')\Phi_{i}(t)}/2$. Hence, one can obtain the steady state $\mathcal{V}$ by solving the following Lyapunov equation as
\begin{equation}
	\label{10}
	\Gamma\mathcal{V}+\mathcal{V}\Gamma^T=-\Lambda,
\end{equation}
where $\Lambda$ is the diffusion matrix defined by
$\Lambda_{ij}\delta(t'-t)=\braket{n_i(t)n_j(t')+n_j(t')n_i(t)}/2 $. It can be rewritten as 
$\Lambda=\text{diag}[\kappa_a(2\bar{n}_a+1),\kappa_a(2\bar{n}_a+1),\kappa_m(2\bar{n}_m+1),\kappa_m(2\bar{n}_m+1),0,\gamma_{b}(2\bar{n}_b+1)]$.

\vspace{-0.9em}
\section{nonreciprocity of the entanglement}
\label{sec3}

\begin{figure*}[ht]
	\centering
	\subfigure{\label{A4}\includegraphics[scale=0.29]{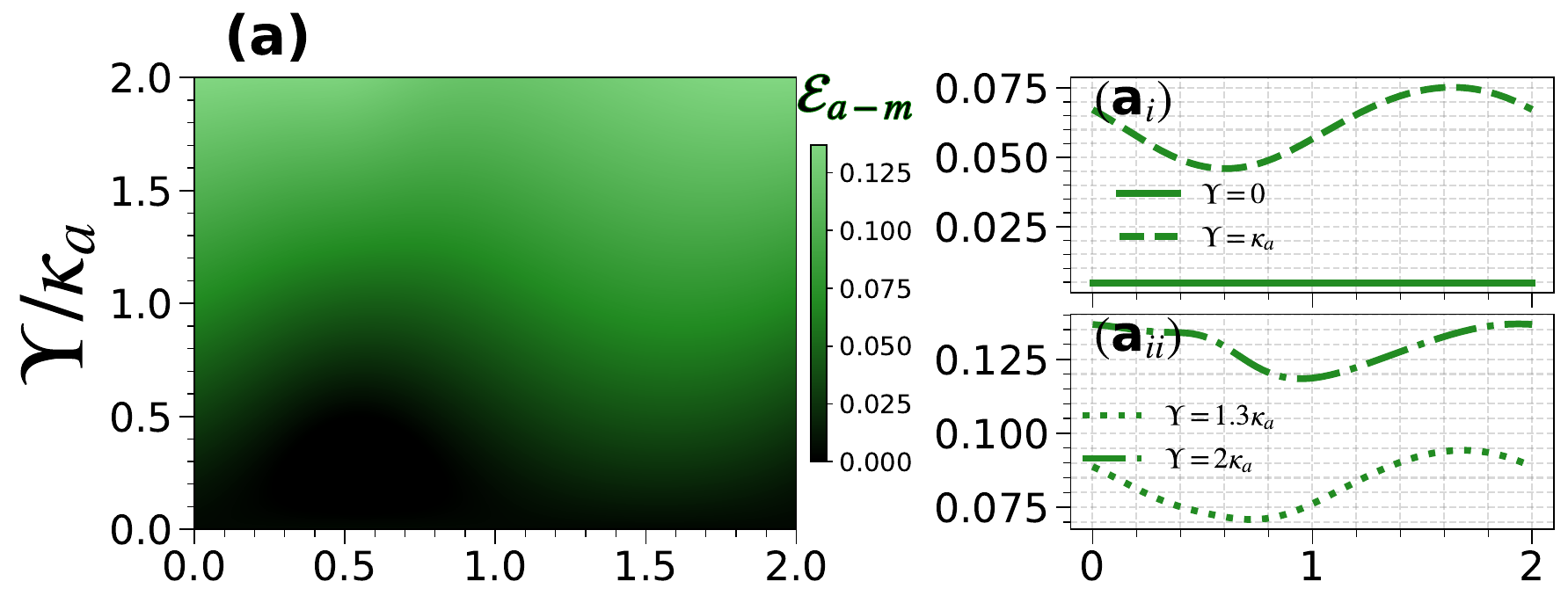}}
	\subfigure{\label{A4}\includegraphics[scale=0.29]{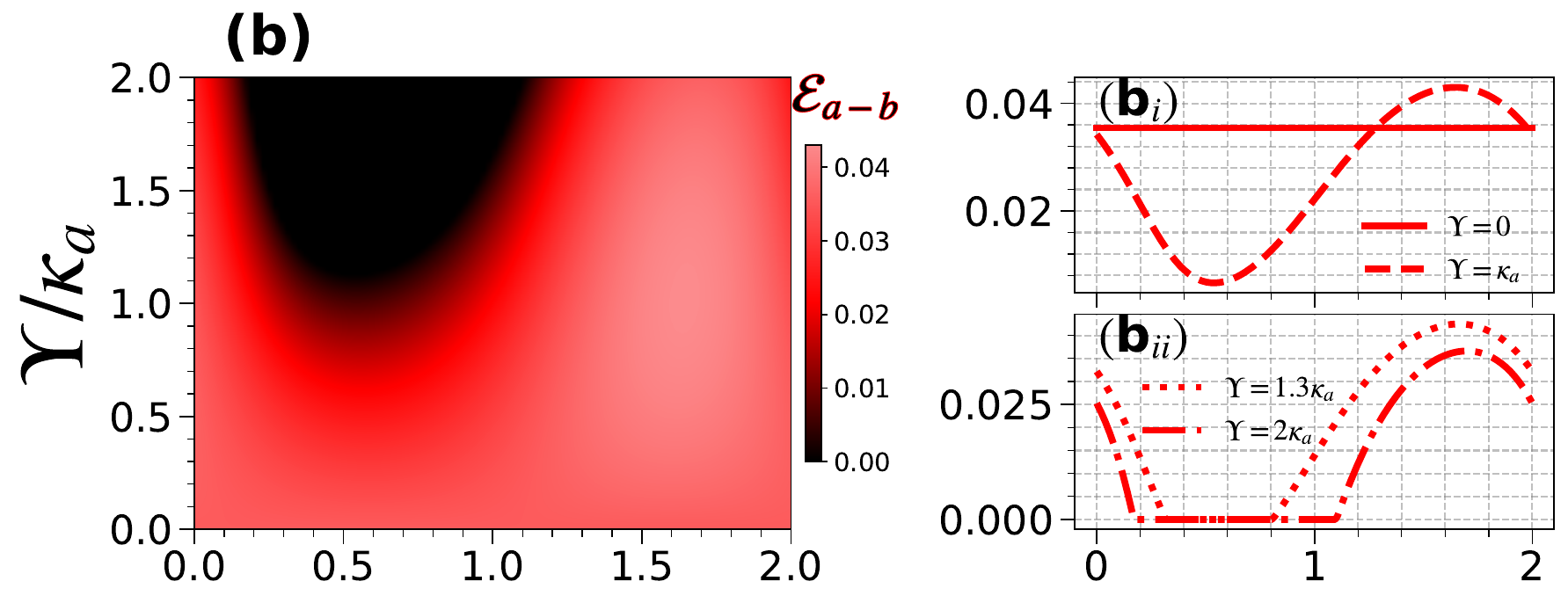}} \\[-1.4em] 
	\hspace{0.4em}	\subfigure{\label{A4}\includegraphics[scale=0.28]{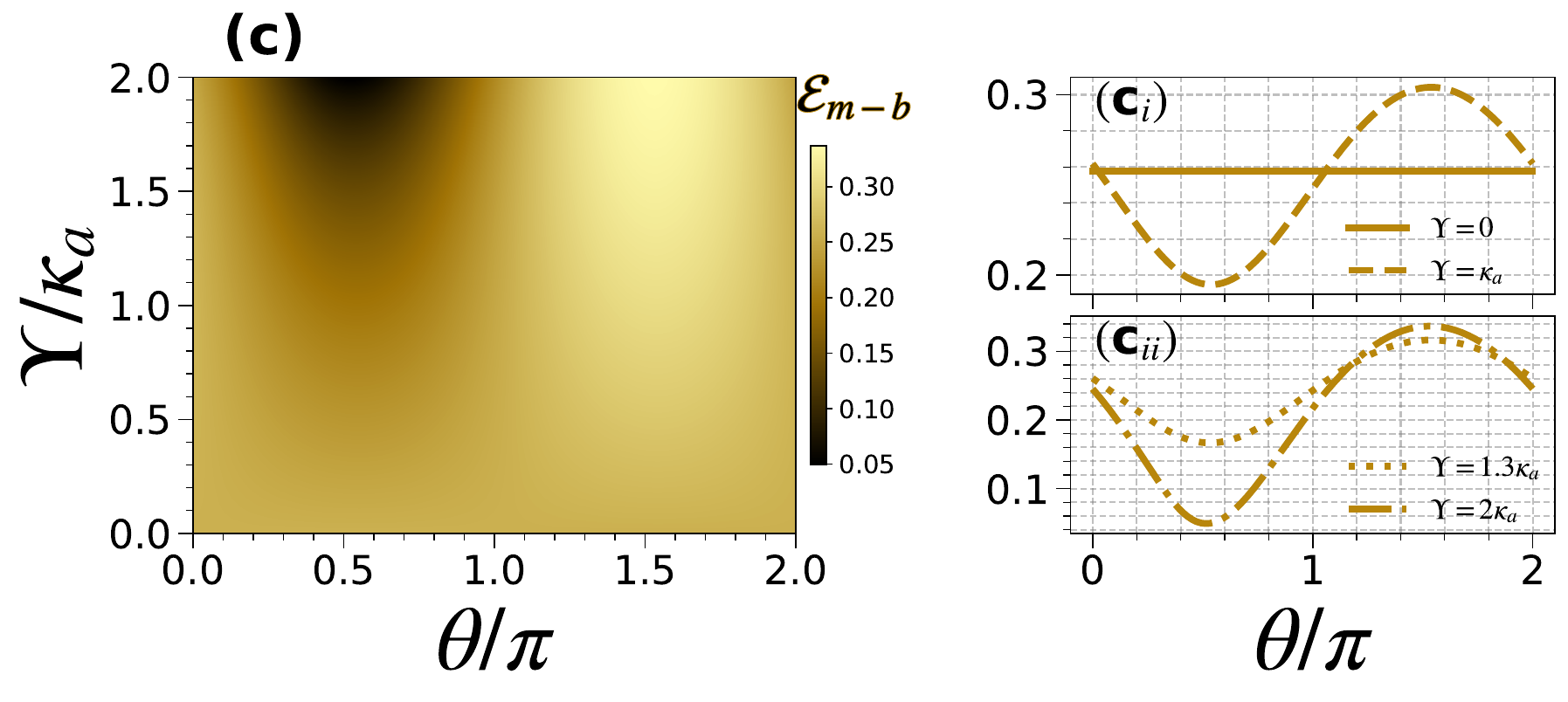}}
	\hspace{0.2em}	\subfigure{\label{A6}\includegraphics[scale=0.28]{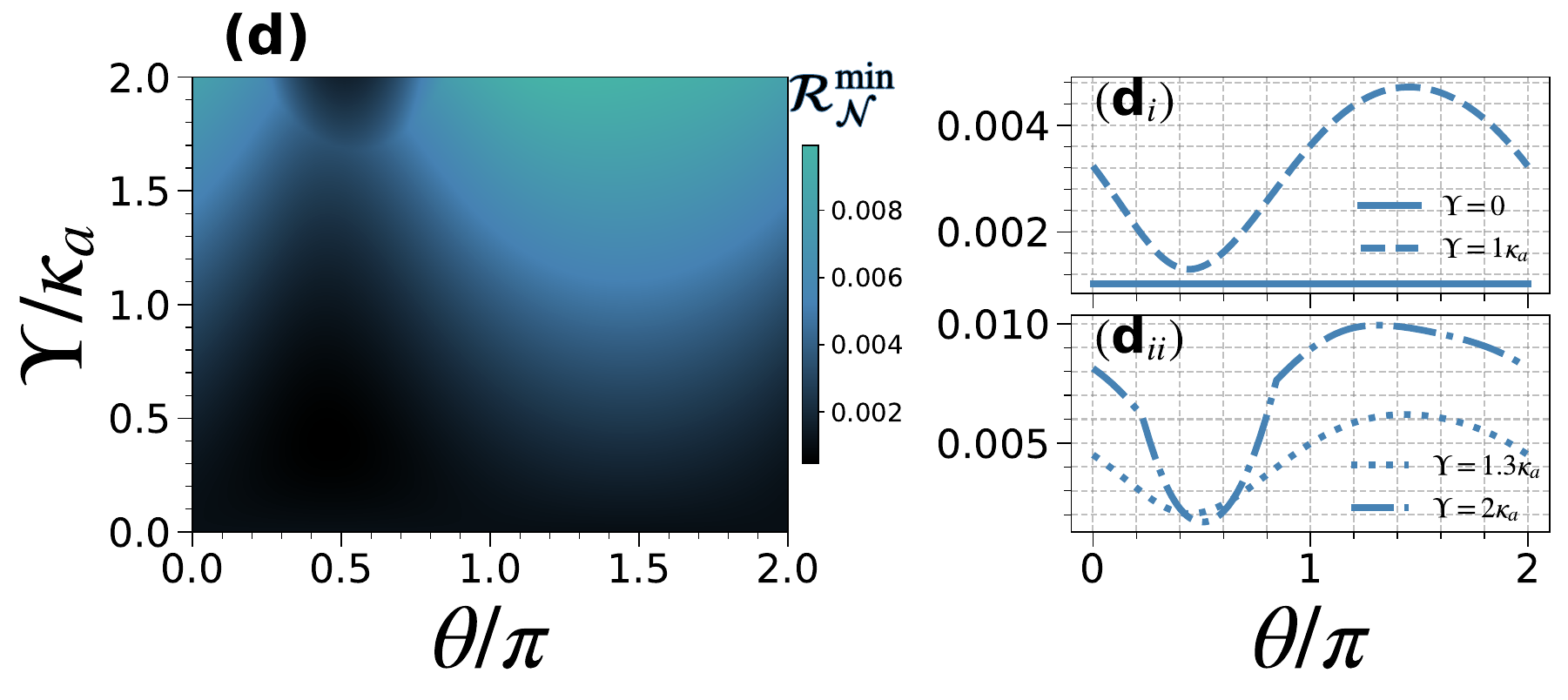}}           
	\captionsetup{justification=RaggedRight, singlelinecheck=false}
	\vspace{-1em} 
	\caption{{\small Density plots of logarithmic negativity: (a) $\mathcal{E}_{a-m}$ (green), (b) $\mathcal{E}_{a-b}$ (red), (c) $\mathcal{E}_{m-b}$ (brown), and (d) the minimum residual contangle $\mathcal{R}_\mathcal{N}^{\min}$ (blue), versus the squeezed parameter $\Upsilon$ and phase parameter $\theta$. Moreover, we set $\Upsilon=0$ (solid line) and $\Upsilon=\kappa_{_a}$ (dashed line) in (a$_i$)-(d$_i$). Besides, $\Upsilon=1.3\kappa_{_a}$ (dotted line) and $\Upsilon=2\kappa_{_a}$ (dot-dashed line) in (a$_{ii}$)-(d$_{ii}$). Finally, $\Delta_{a}=\Delta_{m}=\omega_{_b}$, $T=10$ mK. The other parameters are provided in the main text.}}
	\label{fig:2}
\end{figure*}

To quantify the quantum entanglement among the photon-magnon-phonon modes in the system, we adopt the logarithmic negativity $\mathcal{E}_\mathcal{N}$ for bipartite entanglement between the subsystems $\alpha$ and $\beta$ (where $\alpha,\beta \in \{a,m,b\}$ and $\alpha \neq \beta$) as \cite{E5,E6}
\begin{equation}
	\mathcal{E}_{\alpha-\beta} \equiv \max[0, -\ln(2\tilde{\nu}_{\alpha\beta})],
\end{equation}
where $\tilde{\nu}_{\alpha\beta}$ is the minimum symplectic eigenvalue of the partially transposed $4\times4$ covariance submatrix $\tilde{\mathcal{V}}_{\alpha\beta}$.
For tripartite entanglement, a \textit{bona fide} quantifier is given by the \textit{minimum} residual contangle $\mathcal{R}_\mathcal{N}^{\min}$ \cite{E7,E8,E9}, defined as
\begin{equation}
	\mathcal{R}_\mathcal{N}^{\min}\equiv \min\big[\mathcal{R}_\mathcal{N}^{a|mb},\mathcal{R}_\mathcal{N}^{m|ab},\mathcal{R}_\mathcal{N}^{b|am}\big],
\end{equation}
where $\mathcal{R}_\mathcal{N}^{\alpha|\beta \delta }\equiv C_{\alpha|\beta\delta}-C_{\alpha|\beta}-C_{\alpha|\delta}$ denotes the residual contangle, where $C_{i|j}$ being the contangle of the subsystems $i$ and $j$ ($j$ contains one or two modes), defined as the squared logarithmic negativity (see \textbf{Appendix B} for details) \cite{A2}. This satisfies the monogamy of quantum entanglement, namely $C_{i|jk}\ge C_{i|j}+C_{i|k}$. Note that a non-zero minimum residual angle $\mathcal{R}_\mathcal{N}^{\min}>0$ suggests the presence of tripartite entanglement in the system.

Furthermore, to quantify the nonreciprocal behavior of entanglement \cite{C8}, we introduce the bidirectional contrast ratio $\mathcal{C}$ for bipartite and tripartite entanglement as
\begin{eqnarray}
		\label{Non}
	\mathcal{C}_{\mathcal{E}}&=& \frac{|\mathcal{E}_{\mathcal{N}}(\theta)-\mathcal{E}_{\mathcal{N}}(\theta+\pi)|}{\mathcal{E}_{\mathcal{N}}(\theta)+\mathcal{E}_{\mathcal{N}}(\theta+\pi)},\\ \nonumber
	\mathcal{C}_{\mathcal{R}}&=& \frac{|\mathcal{R}_{\mathcal{N}}(\theta)-\mathcal{R}_{\mathcal{N}}(\theta+\pi)|}{\mathcal{R}_{\mathcal{N}}(\theta)+\mathcal{R}_{\mathcal{N}}(\theta+\pi)}.
\end{eqnarray}
Here, $\theta$ and $\theta+\pi$ correspond to opposite squeezing phases, which induce opposite signs for both the frequency shift $(\Delta_{\theta}= \Upsilon \sin\theta)$ and the damping shift $(\kappa_{\theta}= \Upsilon \cos\theta)$ of the magnon mode. 
This notation is analogous to the contrast ratios defined in Refs.~\cite{C6,CCC5,C8,D1,D2}.
In our case, $\theta \in (0,\pi)$ corresponds to $\Delta_\theta>0$, while $\theta \in (\pi,2\pi)$ corresponds to $\Delta_\theta<0$, as illustrated in Table~\ref{tab:1}.
This phase-controlled sign reversal is the main mechanism underlying the nonreciprocal entanglement. The value $\mathcal{C}_{\mathcal{E}} (\mathcal{C}_{\mathcal{R}}) = 1$ represents an ideal nonreciprocity, while $\mathcal{C}_{\mathcal{E}} (\mathcal{C}_{\mathcal{R}}) =0$ indicates a complete absence of nonreciprocity for bipartite (tripartite) entanglement.

\begin{table}[h]
	\captionsetup{justification=raggedright, singlelinecheck=false}
	\caption{Sign intervals of the frequency shift $\Delta_{\theta}$ and damping shift $\kappa_{\theta}$ versus the squeezing phase $\theta$.}
	\vspace{0.3em} 
	\centering
	\begin{tabular}{c c c c}
		\toprule
		\textbf{Sign} & $\boldsymbol{\Delta_{\theta}}$ & $\boldsymbol{\kappa_{\theta}}$ & $\boldsymbol{(\Delta_{\theta}, \kappa_{\theta})}$ \\
		\midrule
		$> 0$ ~~~~
		& $(0,\ \pi)$ 
		&~~$\left( 0,\ \tfrac{\pi}{2} \right) \cup \left( \tfrac{3\pi}{2},\ 2\pi \right)$ ~~~
		& $\left( 0,\ \tfrac{\pi}{2} \right)$ \\[0.5em]
		$< 0$ ~~~~
		& $(\pi,\ 2\pi)$ 
		&~~$\left( \tfrac{\pi}{2},\ \tfrac{3\pi}{2} \right)$~~
		& $\left( \pi,\ \tfrac{3\pi}{2} \right)$ \\
		\bottomrule
	\end{tabular}
	\label{tab:1}	
\end{table}

The physical mechanism behind nonreciprocal entanglement in our system differs significantly from conventional approaches. While methods based on Kerr, Sagnac, or Barnett effects generally produce frequency (red or blue) shifts alone \cite{C6,CCC5,C8,D1,D2}, magnon squeezing generates a richer structure through dual control.
This mechanism simultaneously produces effective frequency shifts that depend on the amplitude and phase parameters of the squeezing process, while also modifying the magnon decay rate. Note that the results presented here are obtained using experimentally feasible parameters \cite{A01,A1,A2}, namely $\omega_{m}/2\pi=10$ GHz, $\omega_b/2\pi=10$ MHz, $\kappa_a/2\pi=3$ MHz, $\kappa_{m}=\kappa_a/5$, $G_{m}/2\pi=g_{a}/2\pi=4.8$ MHz, $\gamma_{b}/2\pi=10^2$ Hz, and the temperature $T = 10$ mK (the lowest temperature of a dilution refrigerator). In this situation, $\Delta_a\bar{\Delta}_m\simeq\omega_{_b}^2\gg g_{_a}^2$, where the effective magnomechanical coupling is $G_{_m}\simeq \sqrt{2}g_{m}(\omega_b+i\Upsilon e^{i\theta})/(\omega_{_b}^2-\Upsilon^2)$ (see Eq.~(\ref{04})). Indeed, $G_{_m}=4.8$ MHz can be achieved by using a driving magnetic field $H_d\simeq 8.11\times 10^{-5}$ T with $\theta=\pi/2$ ($H_d\simeq 3.65\times 10^{-5}$ T with $\theta=3\pi/2$), $g_{_m}=0.2$ Hz \cite{F00}, and $\Upsilon=1.3\kappa_{_a}$. The results presented are obtained when the system is stable, in accordance with the Routh-Hurwitz criterion \cite{F0}.

To achieve magnon squeezing and observe significant nonreciprocal entanglement, it is essential to drive the magnon mode with a red-detuned microwave field $\Delta_m\simeq\omega_{_b}$, as it effectively cools the vibrational mode close to its quantum ground state. The sign of $\Delta_{\theta}$ plays a crucial role in determining the nonreciprocal entanglement in the system. This approach presents several advantages with respect to the other methods. It allows for controlled variation in detuning by modifying the phase parameter, which gives access to tunable entanglement and enables nonreciprocal behavior. The synergy between phase-dependent detuning ($\pm\Delta_{\theta}$) and dissipation control ($\pm\kappa_{{\theta}}$) enables directional entanglement enhancement, a unique method unattainable with other approaches \cite{C6,CCC5,C8,D1,D2}, offering flexibility in the choice of directional control. Given that $\Upsilon$ is non-negative, the sign of $\Delta_{\theta}$ and $\kappa_{\theta}$ depends on the phase parameter as indicated in Table~.\ref{tab:1}. Under the phase transformation $\theta \to \theta + \pi$, the quantities $\Delta_{\theta}$ and $\kappa_{\theta}$ reverse their signs, i.e., $\Delta_{\theta}\to -\Delta_\theta$ and $\kappa_{\theta}\to -\kappa_{\theta}$. This transformation produces a directional dynamics that gives rise to the nonreciprocal entanglement.

\begin{figure}[t]	
	\begin{center}
		\subfigure{\label{A4}\includegraphics[scale=0.25]{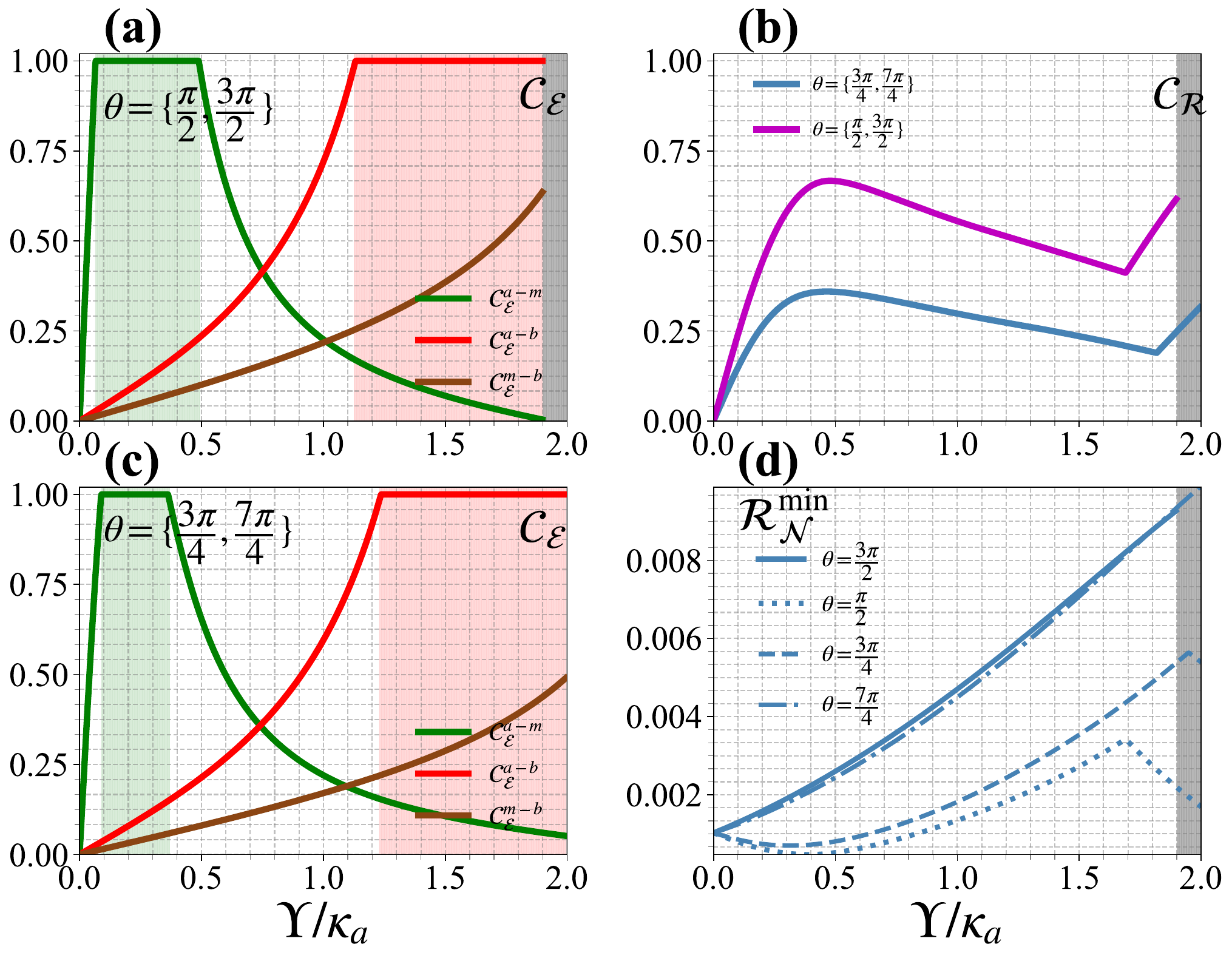}}
	\end{center}
	\captionsetup{justification=RaggedRight, singlelinecheck=false}
	\vspace{-1.5em} 
		\caption{{\small Bidirectional contrast ratios $\mathcal{C}_{\mathcal{E}}^{a-m}$ (green), $\mathcal{C}_{\mathcal{E}}^{a-b}$ (red), and $\mathcal{C}_{\mathcal{E}}^{m-b}$ (brown) versus the squeezing parameter $\Upsilon$ for (a) $\theta=\{\pi/2,3\pi/2\}$ and (c) $\theta=\{3\pi/4, 7\pi/4\}$. (b) $\mathcal{C}_{\mathcal{R}}$ versus the squeezing parameter $\Upsilon$ for $\theta=\{\pi/2,3\pi/2\}$ (purple) and $\theta=\{3\pi/4, 7\pi/4\}$ (blue). (d) The minimum residual contangle $\mathcal{R}_\mathcal{N}^{\min}$ versus the squeezing parameter $\Upsilon$. The colored vertical stripes represent the ideal nonreciprocal entanglement zones ($\mathcal{C}_{\mathcal{E}}=1$). The gray areas denote the regimes where the system is unstable. The other parameters are the same as in Fig.\ref{fig:2}.}}
	\label{fig:3}
\end{figure}

Fig.~\ref{fig:2}(a) illustrates how the entanglement $\mathcal{E}_{a-m}$ changes during the magnon squeezing process with respect to the parameters $\theta$ and $\Upsilon$. The entanglement increases progressively with the squeezing amplitude, reaching its maximum for $\Upsilon = 1.3\kappa_a$, confirming the efficiency of the squeezing process.
As shown in Fig.~\ref{fig:2}(b), there is a clear decrease in the entanglement $\mathcal{E}_{a-b}$ with a maximum amplitude of $0.04$. The non-uniform spatial behavior shows maximum values at certain points and zero at symmetric points, revealing nonreciprocal behavior. Moreover, for $\Upsilon =1.3\kappa_a$, the asymmetry becomes evident, with a maximum at $\theta = 3\pi/2$ and vanishing entanglement at $\theta= \pi/2$. 
In contrast, as illustrated in Fig.~\ref{fig:2}(c), the entanglement $\mathcal{E}_{m-b}$ demonstrates a significant enhancement, reaching a value of $0.33$. Despite maintaining the same spatial behavior as Fig.~\ref{fig:2}(b), the amplitude is significantly amplified, preserving the nonreciprocal properties with enhanced efficiency. Finally, Fig.~\ref{fig:2}(d) shows the tripartite entanglement of the system $\mathcal{R}_\mathcal{N}^{\min}$ with amplitude $0.05$. As with the previous cases, this configuration reproduces the observed amplitude variations, but in a multipartite context, offering a complementary perspective on the quantum correlations of the MM cavity. Remarkably, at these phase values $\theta=\pi/2$ and $\theta=3\pi/2$, the dissipation control vanishes ($\kappa_{\theta}=0$), while the frequency shift reaches its maximum magnitude ($|\Delta_{\theta}|=\Upsilon$), isolating the only frequency-shift mechanism analogous to conventional approaches based solely on frequency control such as the Kerr effect or Barnett effect \cite{C6,CCC5,C8,D1,D2}.

\begin{figure}[t]
	\centering
	\subfigure{\label{A4}\includegraphics[scale=0.3]{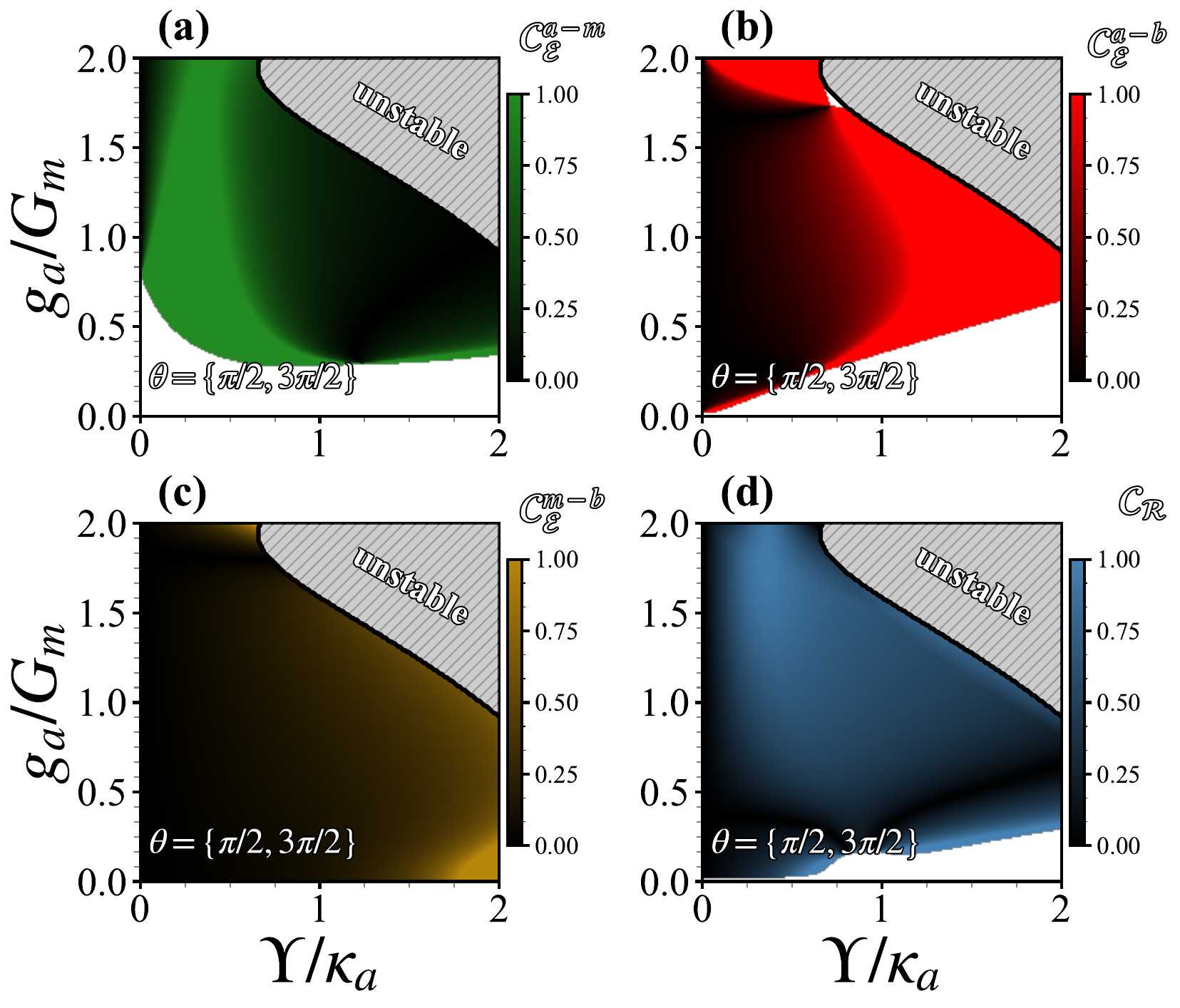}}
	\captionsetup{justification=RaggedRight, singlelinecheck=false}
	\vspace{-1.5em} 
	\caption{{\small Density plots of bidirectional contrast ratios (a) $\mathcal{C}_{\mathcal{E}}^{a-m}$, (b) $\mathcal{C}_{\mathcal{E}}^{a-b}$, (c) $\mathcal{C}_{\mathcal{E}}^{m-b}$, and (d) $\mathcal{C}_{\mathcal{R}}$, versus the cavity-magnon coupling $g_{_a}$ and the squeezed parameter $\Upsilon$ for $\theta=\{\pi/2,3\pi/2\}$. The gray area indicates the regime where the system is unstable. The white area reflects the regime where entanglement disappears in both directional configurations. The other parameters are the same as in Fig.\ref{fig:2}.}}
	\label{fig:4}
\end{figure}

Now, we quantify the nonreciprocal entanglement behavior using contrast ratios via comparing pure frequency shift and dual-control configurations. Indeed, Fig.~\ref{fig:3}(a) shows the evolution of nonreciprocity contrast $\mathcal{C}_{\mathcal{E}}$ versus $\Upsilon$ for the phase configuration $\theta = \{\pi/2,3\pi/2\}$. At these phase values, the dissipation control vanishes ($\kappa_{\theta}=0$), while the frequency shift is maximized ($|\Delta_{\theta}|=\Upsilon$). As shown, $\mathcal{C}_{\mathcal{E}}^{a-m}$ (green) maintains ideal nonreciprocity ($\mathcal{C}_{\mathcal{E}}=1$) when $\Upsilon <0.5\kappa_a$, revealing sensitivity to the squeezing process. In contrast, $\mathcal{C}_{\mathcal{E}}^{a-b}$ (red) reaches an ideal nonreciprocity of entanglement for $\Upsilon$ ranging from $1.12\kappa_{a}$ to $1.19\kappa_{a}$, beyond which the system becomes unstable according to the Routh-Hurwitz criterion \cite{F0}, restricting stable operation to this narrow range.
Fig.~\ref{fig:3}(b) compares the tripartite entanglement nonreciprocity $\mathcal{C}_{\mathcal{R}}$ for two phase configurations. For $\theta =\{\pi/2,3\pi/2\}$ (magenta), where only frequency shift is active, $\mathcal{C}_{\mathcal{R}}$ reaches a maximum of approximately $0.65$. In contrast, for $\theta =\{3\pi/4,7\pi/4\}$ (blue), where both frequency shift ($|\Delta_{\theta}|=\Upsilon/\sqrt{2}$) and dissipation control ($|\kappa_{\theta}|=\Upsilon/\sqrt{2}$) are simultaneously active, $\mathcal{C}_\mathcal{R}$ exhibits reduced amplitude with a maximum of approximately $0.45$, but maintains stability over an extended parameter range, demonstrating the robustness provided by dual control.
Fig.~\ref{fig:3}(c) demonstrates the dual-control configuration $\theta = \{3\pi/4, 7\pi/4\}$. The nonreciprocity contrasts exhibit similar trends to Fig.~\ref{fig:3}(a), with $\mathcal{C}_{\mathcal{E}}^{a-m}$ (green) achieving unity for weak squeezing and $\mathcal{C}_{\mathcal{E}}^{a-b}$ (red) approaching unity for strong squeezing. However, the stable parameter range extends significantly broader than the pure frequency-shift case, confirming that the interplay between frequency and dissipation control provides enhanced robustness against parameter variations, a key advantage of our dual control mechanism over conventional single-parameter approaches.
Finally, Fig.~\ref{fig:3}(d) shows the minimum residual contangle $\mathcal{R}_\mathcal{N}^{\min}$ evolution with the squeezing parameter for different phase configurations. All curves start from similar initial values and diverge as $\Upsilon$ increases, with $\theta = 3\pi/2$ (solid line) exhibiting the strongest enhancement. 
The comparison between $\theta = \pi/2$ (dashed) and $\theta = 3\pi/4$ (dash-dot) demonstrates phase-dependent behavior, confirming the decisive influence of the squeezing phase on tripartite quantum correlations. These results show that the dual-control configuration $\theta =\{3\pi/4,7\pi/4\}$ optimizes the balance between frequency shift and dissipation control, improving stability compared to the pure frequency shift case.

\begin{figure}[t]
	\centering
	\subfigure{\label{A6}\includegraphics[scale=0.3]{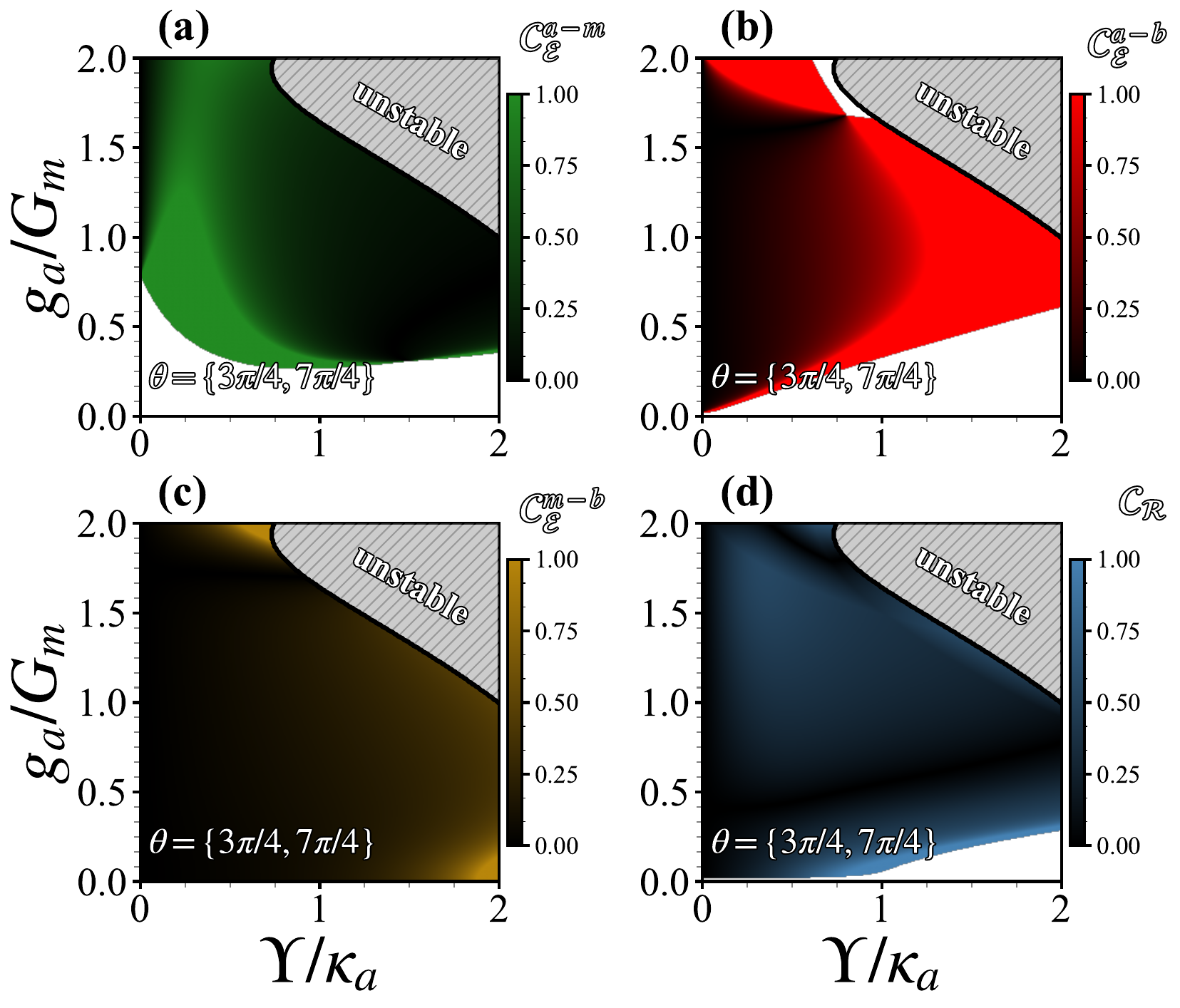}}  
	\captionsetup{justification=RaggedRight, singlelinecheck=false}
	\vspace{-1.5em} 
	\caption{{\small Density plots of bidirectional contrast ratios (a) $\mathcal{C}_{\mathcal{E}}^{a-m}$, (b) $\mathcal{C}_{\mathcal{E}}^{a-b}$, (c) $\mathcal{C}_{\mathcal{E}}^{m-b}$, and (d) $\mathcal{C}_{\mathcal{R}}$, versus the cavity-magnon coupling $g_{_a}$ and the squeezed parameter $\Upsilon$ for $\theta=\{3\pi/4, 7\pi/4\}$. The gray area indicates the regime where the system is unstable. The white area reflects the regime where entanglement disappears in both directional configurations. The other parameters are the same as in Fig.\ref{fig:2}.}}
	\label{fig:5}
\end{figure}

Fig.~\ref{fig:4} illustrates the impact of two key parameters on nonreciprocal entanglement in the proposed cavity magnomechanics system. These parameters are the cavity-magnon coupling strength $g_{_a}$ and the squeezing parameter $\Upsilon$. The results for the phase configuration $\theta= \{\pi/2,3\pi/2\}$ are presented in Fig.~\ref{fig:4}. As illustrated in Fig.~\ref{fig:4}(a), the nonreciprocal entanglement $\mathcal{C}_{\mathcal{E}}^{a-m}$ attains its maximum values with distinct limits, suggesting that robust cavity-magnon couplings $g_{_a}>G_{_m}$ and weak squeezing parameters promote ideal nonreciprocity. 
In addition, Fig.~\ref{fig:4}(b) shows that $\mathcal{C}_{\mathcal{E}}^{a-b}$ demonstrates notable nonreciprocity in specific regions, particularly when $g_{_a}=G_{_m}$. In this case, the nonreciprocity increases with increasing squeezing parameter until reaching ideal nonreciprocity. In Fig.~\ref{fig:4}(c), the ideal nonreciprocity $\mathcal{C}_{\mathcal{E}}^{m-b}$ arises for large cavity-magnon couplings $g_{_a}>G_{_m}$ and high squeezing parameters. Additionally, in Fig.~\ref{fig:4}(d), the nonreciprocal tripartite entanglement $\mathcal{C}_{\mathcal{R}}$ shows clear parameter limits and optimal performance characteristics. These analyses reveal precise optimal conditions for maximizing nonreciprocity, requiring fine control of the $g_{_a}$ and $\Upsilon$ parameters to achieve the desired performance. The distinct spatial distributions between the different types of entanglement highlight the versatility of the system in adapting the nonreciprocal behavior to specific coupling configurations. 
Note that the white regions indicate parameters where entanglement vanishes in both directional configurations (according to Eq.~\eqref{Non}), while the colored regions correspond to stable parameter space.
As shown in Fig.~\ref{fig:5}, the alternative squeezing phase $\theta= \{3\pi/4, 7\pi/4\}$ exhibits a near-similar behavior to the previous configuration, with both frequency shift and dissipation control simultaneously active.
Figs.~\ref{fig:5}(a) and (b) illustrate ideal nonreciprocity regions repositioned towards intermediate values $g_a$, demonstrating the crucial influence of the squeezing phase on optimal zone location.
This dual control configuration exhibits enhanced stability compared to the pure frequency-shift case, as evidenced by reduced unstable regions.
Fig.~\ref{fig:5} (c) and (d) show small ideal nonreciprocity regions compared to $\theta= \{\pi/2,3\pi/2\}$.
However, a key advantage of the $\theta= \{3\pi/4, 7\pi/4\}$ configuration is that it provides ideal nonreciprocity with robustness against parameter variations.
The magnetostrictive coupling creates distinct threshold effects, where ideal nonreciprocity only appears above critical coupling strengths and squeezing parameters.

\begin{figure}[t]	
	\begin{center}
		\subfigure{\label{A4}\includegraphics[scale=0.25]{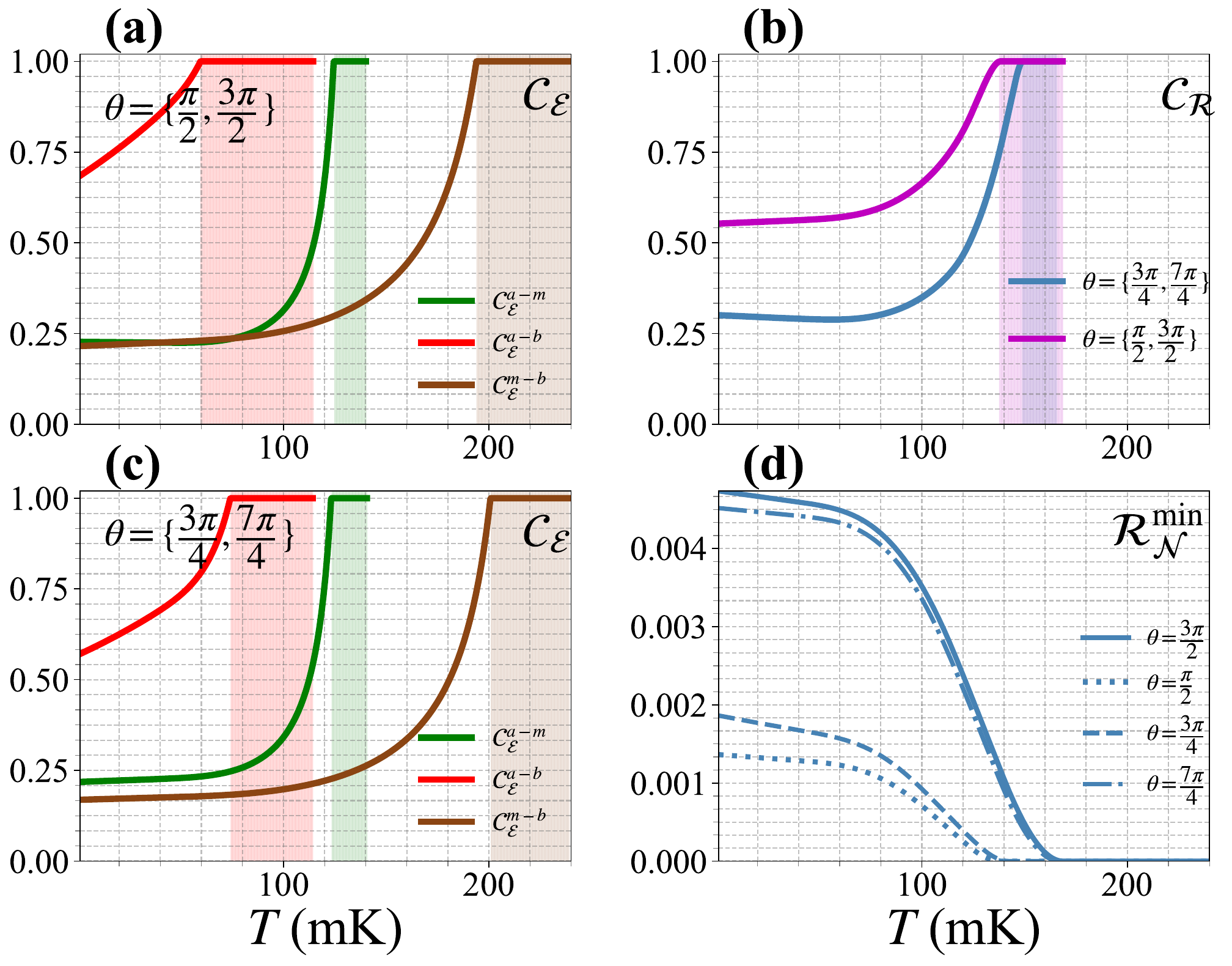}}
	\end{center}
	\captionsetup{justification=RaggedRight, singlelinecheck=false}
	\vspace{-1.3em} 
	\caption{{\small Bidirectional contrast ratios $\mathcal{C}_{\mathcal{E}}^{a-m}$ (green), $\mathcal{C}_{\mathcal{E}}^{a-b}$ (red), and $\mathcal{C}_{\mathcal{E}}^{m-b}$ (brown) versus the bath temperature $T $(mK) for (a) $\theta=\{\pi/2,3\pi/2\}$ and (c) $\theta=\{3\pi/4, 7\pi/4\}$. (b) $\mathcal{C}_{\mathcal{R}}$ versus $T $(mK) for $\theta=\{\pi/2,3\pi/2\}$ (purple) and $\theta=\{3\pi/4, 7\pi/4\}$ (blue). (d) $\mathcal{R}_\mathcal{N}^{\min}$ versus the bath temperature $T $(mK). The coloured vertical stripes represent the ideal nonreciprocal entanglement zones. The parameters are the same as in Fig.\ref{fig:2}.}}
	\label{fig:6}
\end{figure}

Fig.~\ref{fig:6} illustrates the temperature dependence of the nonreciprocity across different phase configurations.
Figs.~\ref{fig:6}(a-c) show the nonreciprocity of bipartite entanglement $\mathcal{C}_\mathcal{E}$ and tripartite entanglement $\mathcal{C}_\mathcal{R}$ versus temperature for the phase configurations $ \theta= \{\pi/2,3\pi/2\}$, and $\theta= \{3\pi/4, 7\pi/4\}$.
Fig.~\ref{fig:6}(a) corresponds to the phase configuration $\theta = \{\pi/2,3\pi/2\}$ and reveals ideal nonreciprocity at temperatures ranging from $60$mK-$115$mK for $\mathcal{C}_{\mathcal{E}}^{a-b}$, $125$mK-$140$mK for $\mathcal{C}_{\mathcal{E}}^{a-m}$, and $195$mK-$240$mK for $\mathcal{C}_{\mathcal{E}}^{m-b}$.
Notably, $\mathcal{C}_{\mathcal{E}}^{m-b}$ supports ideal nonreciprocity at the highest temperatures despite the large phonon thermal occupation $\bar{n}_b \simeq  k_B T/(\hbar \omega_b) \gg 1$ at megahertz frequencies. This can be attributed to the red-detuned regime $\Delta_a = \bar{\Delta}_m \simeq \omega_b$, which cools the mechanical mode effectively,  sustaining magnon-phonon correlations \cite{A2}.
Fig.~\ref{fig:6}(b) compares the tripartite entanglement nonreciprocity $\mathcal{C}_{\mathcal{R}}$ for two phase configurations, with $\mathcal{C}_{\mathcal{R}}$ (blue) for $\theta = \{3\pi/4, 7\pi/4\}$ exhibiting ideal nonreciprocity in the range $150$ mK--$166$ mK, while $\mathcal{C}_{\mathcal{R}}$ (purple) for $\theta = \{\pi/2,3\pi/2\}$ exhibits ideal nonreciprocity from $140$ mK to $170$ mK. This comparison highlights the distinct temperature thresholds above which nonreciprocity disappears.
Fig.~\ref{fig:6}(c), which corresponds to the phase configuration $\theta = \{3\pi/4, 7\pi/4\}$, shows ideal nonreciprocity over temperature ranges of $75$ mK-$114$ mK for $\mathcal{C}_{\mathcal{E}}^{a-b}$, $123$ mK-$140$ mK for $\mathcal{C}_{\mathcal{E}}^{a-m}$, and $201$ mK-$240$ mK for $\mathcal{C}_{\mathcal{E}}^{m-b}$.
The same thermal ordering is observed, confirming its robustness across both phase configurations.
Fig.~\ref{fig:6}(d) shows the minimum residual contangle $R_\mathcal{N}^{min}$, which decreases monotonically with increasing temperature for all phase configurations (represented by different line styles), indicating degradation of the quantum correlations with increasing temperature. For both configurations, $R_\mathcal{N}^{min}$ is larger at the negative frequency shift phase ($\Delta_{\theta}<0$), i.e., $\theta = 3\pi/2$ over $\pi/2$ and $\theta = 7\pi/4$ over $3\pi/4$.

Note that the validity of all the above figures depends on the condition that the number of magnonic excitations $\braket{m^\dagger m}\ll2N_0s=5N_0$, where $s=\frac{5}{2}$ as mentioned in sec.\ref{sec2}. For a $250\mu$m-diameter YIG sphere, the number of spins $N_0=3.5\times 10^{16}$, 
$|\braket{m}| \simeq1.69\times 10^7$ corresponds to the MM coupling $G_{m}/2\pi=4.8$MHz, $\theta=\pi/2 $ and Rabi frequency $\Omega_0\simeq 1.48 \times10^{15}$Hz which corresponds to the drive magnetic field $H_d \simeq 2.87 \times 10^{-5}\,\text{T}$, and drive power $\mathcal{P}_{\text{m}}=38.5$mW \cite{Exp1,Exp2}. This leads to $|\braket{m^\dagger m}| \simeq  2.88 \times 10^{14}\ll 5N_0=1.8\times10^{17}$. Thus, the condition is satisfied.
However, high pumping power is adopted for magnon modes, which can introduce unwanted nonlinearity due to the nonlinear Kerr term $\mathcal{K}m^{\dagger}mm^{\dagger}m$ in the Hamiltonian \cite{X1,D8}.
To ensure that the $\mathcal{K}$ remains negligible, the $\mathcal{K}|\braket{m}|^3\ll\Omega_{0}$ condition must be satisfied. The positive Kerr coefficient varies inversely proportional to the sphere volume, for which we use a $250\mu$m-diameter YIG sphere, $\mathcal{K}/2\pi=6.4$nHz.
With the parameters used in the paper, $\mathcal{K}|\braket{m}|^3= 1.96 \times 10^{14}$Hz$\ll\Omega_{0}$, validating that linearization provides a good approximation. Hence, the nonlinear effects remain negligible.

\begin{figure}[t]	
	\begin{center}
		\subfigure{\label{A4}\includegraphics[scale=0.25]{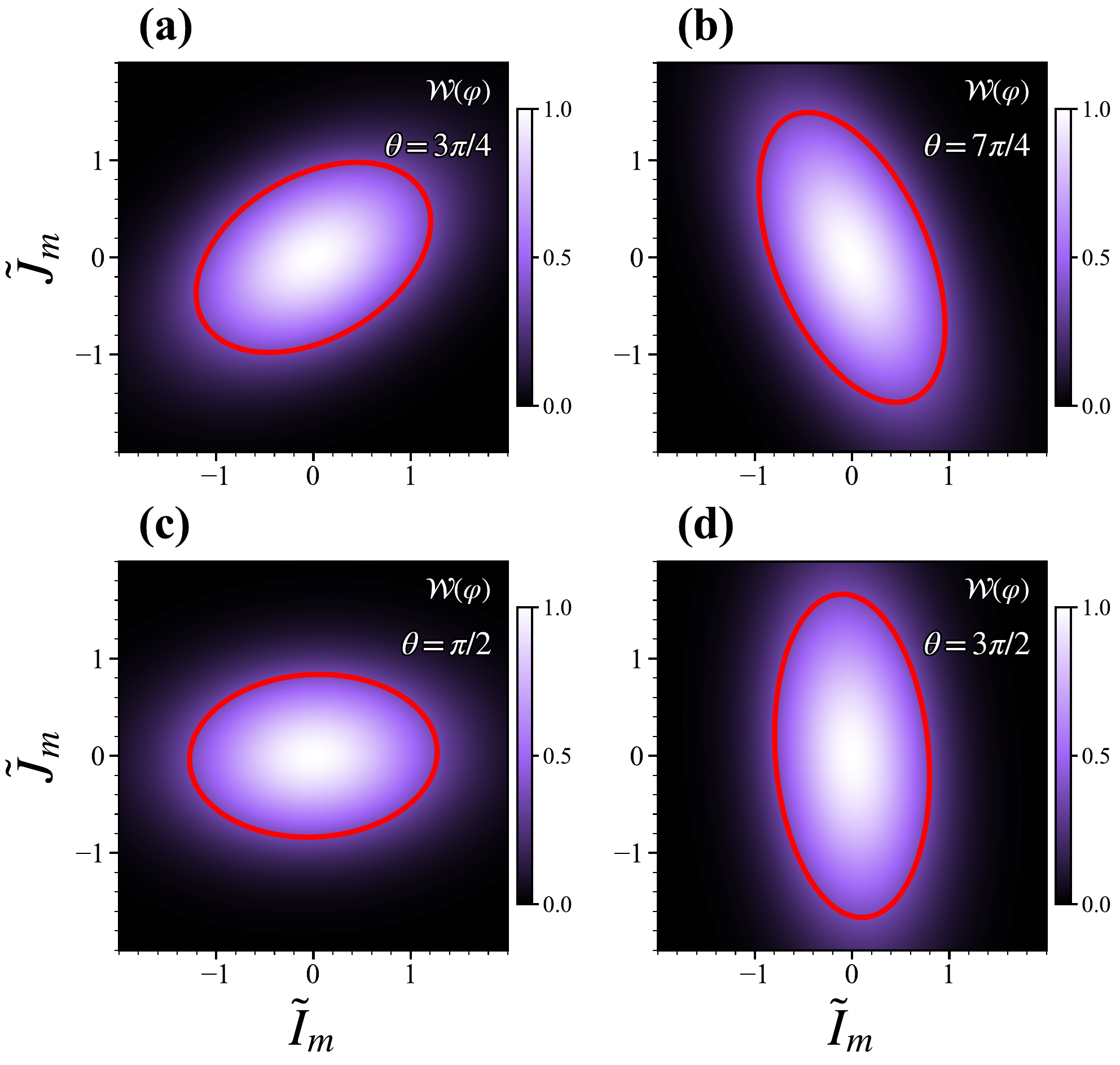}}
	\end{center}
	\captionsetup{justification=RaggedRight, singlelinecheck=false}
	\vspace{-1.3em} 
	\caption{{\small Density plots of the reconstructed Wigner function $\mathcal{W}(\varphi)$ for quadrature magnon mode pairs at different values of squeezing phase: (a) $\theta=3\pi/4$, (b) $\theta=7\pi/4$, (c) $\theta=\pi/2$, and (d) $\theta=3\pi/2$. Solid red contours indicate the $1/e$ of the maximum value of $\mathcal{W}(\varphi)$. These configurations are used to obtain nonreciprocal entanglement with parameters $\Delta_a=\Delta_m$ $=\omega_{_b}, \Upsilon=1.3\kappa_{_a}$, $T=10$mK, and other parameters are listed in the main text.}}
	\label{fig:7}
\end{figure}

Finally, to verify the presence of magnon squeezing, we examine the intrinsic phase-space fluctuations of the magnon mode through its quadrature operators $\{\tilde{I}_m,\tilde{J}_m\}$. We reconstruct the Wigner function $\mathcal{W}(\varphi)$ for the magnon mode, which provides a full representation of its Gaussian quantum state and allows direct visualization of the squeezing phenomenon. The Wigner function is expressed as 

\begin{equation}
	\mathcal{W}(\varphi)= \frac{\exp(-(\varphi\tilde{\mathcal{V}}^{-1}\varphi^\dagger)/2)}{\pi^2\sqrt{\det\mathcal{V}}},
\end{equation}
where $\varphi$ is the magnon quadrature vector and $\tilde{\mathcal{V}}$ is the $2\times2$ magnon sub-covariance matrix extracted from $\mathcal{V}$ \cite{AS1}. Fig.~\ref{fig:7} displays the Wigner function at different squeezing phases, confirming \textit{genuine} magnon squeezing in the parameter regimes used to achieve nonreciprocal entanglement in Figs.~\ref{fig:2}--\ref{fig:6}.
The Wigner functions exhibit elliptical distortion with solid red contours (at $1/e$ of the maximum Wigner function value) showing compression along one quadrature, which is the signature of squeezing. The phase-dependent orientation of the squeezing axis directly visualizes the underlying mechanism where varying $\theta$ controls both the frequency shift $\Delta_\theta=\Upsilon\sin\theta$ and dissipation rate $\kappa_{\theta}= \Upsilon\cos\theta$, enabling directional control of quantum correlations. 
These Wigner reconstructions confirm that a \textit{genuine} magnon squeezing regime (the physical resource underlying our nonreciprocal entanglement) is present for all phase configurations considered in this study.

\section{Conclusion}
\label{sec4}

In conclusion, we have presented a fundamentally distinct approach to generate and control bipartite and tripartite nonreciprocal macroscopic entanglement in a cavity magnomechanics using squeezed magnons. In contrast to the previous conventional effect-based methods \cite{C6,CCC5,C8,D1,D2}, our approach exploits dual control through phase-dependent frequency shifts and tunable dissipation used to achieve nonreciprocal entanglement via the amplitude and phase control of magnon squeezing. Crucially, all parameters used are experimentally accessible with current cavity magnomechanical platforms. Our results demonstrate that the squeezing phase enables switching of the nonreciprocal behavior, with the transformation $\theta \to \theta + \pi$ reversing both $\Delta_\theta$ and $\kappa_\theta$, offering precise control of nonreciprocal entanglement. We have shown that this approach enables ideal nonreciprocity under the optimal conditions for the cavity-magnon coupling and the bath temperature, with enhanced stability compared to the pure frequency-shift methods. Importantly, we verified the presence of squeezed magnons for all phase configurations used to achieve nonreciprocity, confirming the quantum validity of our approach. The squeezed magnon states obtained in a massive system constitute genuine macroscopic quantum states. This underlines their potential for probing quantum effects at macroscopic scales \cite{F1}. Therefore, our findings highlight an alternative route for the generation of nonreciprocal quantum entanglement through phase-controlled magnon squeezing, opening promising prospects for quantum correlations in magnon-based hybrid systems.

\section*{Appendix A}
\appendix
\label{appendix:A}

\renewcommand{\theequation}{A\arabic{equation}}
\setcounter{equation}{0}

In this Appendix, we provide the derivation of the rotating-frame Hamiltonian used in the main text. For simplicity, we set $\hbar=1$ throughout the following calculations. The Hamiltonian in the laboratory frame (time-dependent) is given by 
	\begin{eqnarray}
		\label{A1}
		\tilde{\mathbf{H}}&=&\omega_a a^\dagger a +\omega_{m} m^\dagger m+\frac{\omega_{b}}{2}(q^2+p^2)\nonumber\\ 
		&+& g_{m}m^\dagger mq + g_{a} (a^\dagger+ a)(m^\dagger+ m)\nonumber\\
		&+& i\left[\Omega_0m^\dagger e^{-i\omega_{0}t} + \frac{\Upsilon}{2}m^{\dagger2} e^{i(\theta-2\omega_{0}t)} -\text{H.c.}\right],
	\end{eqnarray}
where $\omega_a$, $\omega_m$, and $\omega_b$ are the frequencies of the cavity photon, magnon, and phonon modes, respectively. To eliminate the explicit time dependence, we apply the unitary transformation $\mathcal{U}=\exp[i\omega_0(a^\dagger a + m^\dagger m)t]$ to transform the above Hamiltonian into a rotating frame at the drive frequency $\omega_0$. The frame-rotated Hamiltonian in the interaction picture is:
	\begin{eqnarray}
		\label{A2}
		\mathbf{H} = \mathcal{U}\tilde{\mathbf{H}}\mathcal{U}^\dagger - i\mathcal{U}\partial_t \mathcal{U}^\dagger.
	\end{eqnarray}

Now, we transform each term in Eq.~\eqref{A1} using the Baker-Campbell-Hausdorff (BCH) identity \cite{Rev1}:
	\begin{equation}
		\label{A3}
		e^A B_j e^{-A} = B_j + [A,B_j] + \frac{1}{2}[A,[A,B_j]] + \cdots,
	\end{equation}
where $A = i\omega_0(a^\dagger a + m^\dagger m)t$ and $j$ stands for photon, phonon and magnon modes. The operators are transformed as $\mathcal{U} a \mathcal{U}^\dagger = a e^{-i\omega_0 t}$ and $\mathcal{U} m \mathcal{U}^\dagger = m e^{-i\omega_0 t}$. While the phonon operators remain unchanged since they commute with $a^\dagger a + m^\dagger m$. The terms $\omega_a a^\dagger a$, $\omega_m m^\dagger m$, $\omega_b(q^2+p^2)/2$, and $g_m m^\dagger m q$ are all invariant under this transformation.

Then, applying the rotating-wave approximation to the photon-magnon interaction, $g_a(a^\dagger + a)(m^\dagger + m) \to g_a(a^\dagger m + am^\dagger)$ (valid when $\omega_a,\omega_{m}\gg g_a,\kappa_{a},\kappa_{m}$) \cite{A1,A2}, which also remains invariant since the exponential factors $e^{\pm i\omega_0 t}$ vanish.
For the coherent driving term, the transformation yields to $\Omega_0(m^\dagger e^{i\omega_0 t})e^{-i\omega_0 t} = \Omega_0 m^\dagger$.

For the squeezing term, using the commutation relation $[m^\dagger m, m^{\dagger2}] = 2m^{\dagger2}$, the BCH formula gives $e^{i\omega_0 m^\dagger m t}(m^{\dagger2})e^{-i\omega_0 m^\dagger m t} = m^{\dagger2} e^{2i\omega_0 t}$. Therefore:
	\begin{eqnarray}
		\label{A12}
	e^A \left(\frac{\Upsilon}{2} m^{\dagger2} e^{i(\theta-2\omega_{0}t)}\right) 
e^{-A} = \frac{\Upsilon}{2} e^{i\theta} m^{\dagger2},
	\end{eqnarray}
and similarly for the Hermitian conjugate.

Including the time-derivative contribution $-i\mathcal{U}\partial_t \mathcal{U}^\dagger = -\omega_{0}(a^\dagger a + m^\dagger m)$, the time-independent rotating-frame Hamiltonian becomes:
	\begin{eqnarray}
		\label{A15}
		\mathbf{H}&=&(\omega_a - \omega_0) a^\dagger a +(\omega_m - \omega_0) m^\dagger m+\frac{\omega_{b}}{2}(q^2+p^2)\nonumber\\ 
		&+& g_{m}m^\dagger mq + g_{a} (a^\dagger m+ a m^\dagger) + i[\Omega_0m^\dagger \nonumber\\ 
		&+& \frac{\Upsilon}{2}m^{\dagger2} e^{i\theta} -\text{H.c.}],
	\end{eqnarray}
where $\Delta_a = \omega_a - \omega_0$ and $\Delta_m = \omega_m - \omega_0$ are the detunings, recovering Eq.~\eqref{Hamil} of the main text.\\

Using the rotating-frame Hamiltonian in Eq.~\eqref{A15}, we now describe the system dynamics in the presence of dissipation and environmental noise \cite{Rev02,Rev2,Rev3}. Following the standard treatment of open quantum systems, we account for noise inputs and dissipation in the model. The system evolution is described by the quantum Langevin equations (QLEs). For a general operator variable $O$, the QLEs take the form \cite{A2,Rev4}:
	\begin{eqnarray}
		\label{A16}
		\dot{O} = -i[O, \mathbf{H}] - \zeta O + \aleph,
	\end{eqnarray}
where $\zeta$ represents the dissipation rate, and $\aleph$ is the noise operator. Applying Eq.~\eqref{A16} to the rotating-frame Hamiltonian in Eq.~\eqref{A15}, we derive the standard dynamical equations that include relaxation and noise terms:
	\begin{eqnarray}
		\label{A17}
		\dot{a} &=& -i\Delta_a a- i g_a m  - \kappa_a a + \sqrt{2\kappa_a} a^{\text{in}}, \nonumber\\
		\dot{m} &=& -i\Delta_m m- i g_a a + \Omega_0 - i g_m mq + \Upsilon m^\dagger e^{i\theta} - \kappa_m m \nonumber\\ 
		&+& \sqrt{2\kappa_m} m^{\text{in}}, \nonumber\\
		\dot{q} &=& \omega_b p, \quad \dot{p} = -\omega_b q - g_m m^\dagger m - \gamma_b p + \xi,
	\end{eqnarray}
where $\kappa_a$, $\kappa_m$, and $\gamma_b$ are the decay rates of the cavity, magnon, and phonon modes, respectively. Here, $a^{\text{in}}$, $m^{\text{in}}$, $\xi$ are the corresponding input noise operators, which indicates Eq.~\eqref{2} of the main text.

\section*{Appendix B}
\appendix
\label{appendix:B}

\renewcommand{\theequation}{B\arabic{equation}}
\setcounter{equation}{0}

The bipartite entanglement between subsystems $\alpha$ and $\beta$ ($\alpha,\beta \in \{a,m,b\}$ and $\alpha \neq \beta$) is quantified using the logarithmic negativity \cite{E5,E6}:
	\begin{equation}
		\mathcal{E}_{\alpha-\beta} \equiv \max[0, -\ln(2\tilde{\nu}_{\alpha\beta})],
	\end{equation}
	where $\tilde{\nu}_{\alpha\beta} = \min\left|\text{eig}[i\boldsymbol{\Omega}_2\tilde{\mathcal{V}}_4]\right|$ is the smallest symplectic eigenvalue of the partially transposed covariance matrix, with $\boldsymbol{\Omega}_2 = \bigoplus_{k=1}^{2} i\sigma_y$ being the symplectic matrix and $\sigma_y$ the Pauli-y matrix.
	The partially transposed covariance matrix is given as:
	\begin{equation}
		\tilde{\mathcal{V}}_4 = \mathcal{P}_0 \mathcal{V}_4 \mathcal{P}_0,
	\end{equation}
	where $\mathcal{V}_4$ is the $4 \times 4$ covariance submatrix of the two modes extracted from the full covariance matrix $\mathcal{V}$ in Eq.~(\ref{10}). Here, $\mathcal{P}_0 = \text{diag}(1, -1, 1, 1)$ realizes the partial transposition. 
	For tripartite systems, \textit{genuine} three-mode entanglement is quantified by the minimum residual contangle \cite{E7,E8,E9}:
	\begin{equation}
		\mathcal{R}_\mathcal{N}^{\tau} = \min_{(\alpha,\beta,\delta)} \left\{\mathcal{C}^{\alpha|\beta\delta} - \mathcal{C}^{\alpha|\beta} - \mathcal{C}^{\alpha|\delta}\right\},
	\end{equation}
	where $(\alpha, \beta, \delta)$ denotes all permutations of the three modes $\{a,m,b\}$. The contangle $\mathcal{C}^{i|j}$ is defined as the squared logarithmic negativity \cite{C8}:
	\begin{equation}
		\mathcal{C}^{i|j} \equiv [\mathcal{E}_{i-j}]^2.
	\end{equation}
	For bipartitions where $j$ contains two modes, the symplectic eigenvalue is computed as $\tilde{\nu}_{i|j} = \min\left|\text{eig}[i\boldsymbol{\Omega}_3\tilde{\mathcal{V}}_6]\right|$, with $\boldsymbol{\Omega}_3 = \bigoplus_{k=1}^{3} i\sigma_y$ and $\tilde{\mathcal{V}}_6 = \mathcal{P} \mathcal{V} \mathcal{P}$. The partial transposition matrices for different bipartitions are: 
	\begin{eqnarray}
		\mathcal{P}_{a|mb} &=& \text{diag}(1, -1, 1, 1, 1, 1), \nonumber\\
		\mathcal{P}_{m|ab} &=& \text{diag}(1, 1, 1, -1, 1, 1), \nonumber\\
		\mathcal{P}_{b|am} &=& \text{diag}(1, 1, 1, 1, 1, -1).
\end{eqnarray}

\section*{Acknowledgment}
The authors would like to thank the referees for their report, which helped us to improve our manuscript.

\end{document}